\documentstyle [12pt]{article}
\def\be{\begin{equation}}
\def\ee{\end{equation}}
\def\bea{\begin{eqnarray}}
\def\eea{\end{eqnarray}}

\begin{document}

\begin{titlepage}

\title{ The connection of Monge-Bateman equations with  ordinary
differential 
equations and their generalisation}

\author{ A.N. Leznov\\
{\it  Institute for High Energy Physics, 142284 Protvino,}\\{\it
Moscow Region,
Russia}}
\maketitle

\begin{abstract}

It is shown that the Monge equation is equivalent to the ordinary
differential 
equation $\ddot X=0$ of  free motion. Equations of Monge type  (with
their 
general solutions) are connected with each ordinary differential
equation of  
second order $\ddot X=F(\dot X,X; t)$, integrable by quadratures.
The result is generalised to a system of equations of the second
order,
which is in one to one correspondence  with  the multidimensional 
Monge-Bateman system.

\end{abstract}

\end{titlepage}

\section{Introduction}

The famous Monge equation, which has been quoted  in textbooks for
more than 
150 years, has the form:
$$
\lambda_t=\lambda \lambda_x
$$
Its implicit solution:
$$
x-\lambda t=f(\lambda)
$$
reminds one of the solution of the equation of free motion with
constant 
velocity $\lambda$ with initial value for the coordinate
$x_0=f(\lambda)$ at 
$t=0$.

This similarity is not accidental and in the present note we want to
draw 
attention to the fact of the connection of the Monge equation and the
generalised 
it Bateman equation \cite{bat}:
\be
({\lambda_t\over \lambda_x})_t-({\lambda_t\over \lambda_x})
({\lambda_t\over \lambda_x})_x=0 \label{realbat}
\ee
with the ordinary differential equation $\ddot X=0$. This fact allow
us
to introduce  partial differential equations  connected with each
ordinary 
differential equation of  second order (this limitation is
inessential) which 
possess the same integrable properties as the initial ODE.

This result can be generalised to any system of equations of  second
order,
leading to  a generalisation of the Monge and Bateman type of
equations to the 
multidimensional case. The hydrodynamic system of D.B.Fairlie
\cite{DBF} is the 
simplest example of such a generalisation. 

\section{The main assertion and its proof}

Assertion

Let $\ddot X=F(\dot X,X; t)$ be an ordinary differential equation,
where $F$ 
is an arbitrary function of its arguments and $\dot {}$ denotes
differentiation 
with respect to the independent argument $t$. Then the equation in
partial 
derivatives:
\be
({\lambda_t\over \lambda_X})_t-({\lambda_t\over \lambda_X})
({\lambda_t\over \lambda_X})_X=F(-{\lambda_t\over
\lambda_X},X;t)\label{gbat}
\ee
is exactly integrable (implicitly) simultaneously with the initial
ordinary 
differential equation.

We would like to prove this assertion from both sides. First by using
the known 
solution of the initial ordinary differential equation and secondly by
explicit exchange of variables in it directly.

\subsection{The first proof}

 The general solution of an ordinary differential equation of  second
 order 
depends upon two arbitrary constants $(c^1,c^2)$ which we will
consider as 
functions of two arguments $\lambda (X,t)$.
So we have an implicit definition of the function $\lambda$:
\begin{equation}
X=X(t;c^1(\lambda),c^2(\lambda))\label{1}
\end{equation}
By ${}'$ we denote the derivative of $X$ with respect to the argument
$\lambda$:
$$
X'=X_{c^1}c^1_{\lambda}+X_{c^2}c^2_{\lambda}
$$
It is obvious that dot and prime differentiations are mutually
commutative. We
obtain in consequence after differentiation of (\ref{1}) with respect
to its  independent arguments $X$ and $t$:
$$
1=X'\lambda_X\equiv (X_{c_1}c'_1+X_{c_2}c'_2)\lambda_X,\quad
0=X_t+X'\lambda_t
$$
or
$$
X_t=-{\lambda_t\over \lambda_X}
$$
Differentiation of the last equation once more by the arguments $X,t$
leads to 
the result:
\begin{equation}
X_{tt}+X'_t\lambda_t=-({\lambda_t\over \lambda_X})_t,\quad
X'_t\lambda_X=-({\lambda_t\over \lambda_X})_X\label{2}
\end{equation}
Eliminating $X'_t$ from the last two equalities  and taking into
account that
under differentiation $X_{tt}$ $c^1,c^2$ remain constant  we arrive at
the 
generalised Bateman equation (\ref{gbat}). This way from a given
solution to 
equation it satisfied is decribed in each manual.

The generalized Monge equation may be obtain from (\ref{gbat}) after
making the
identification $ {\lambda_t\over \lambda_X}\to \lambda$ and takes the
form:
\be
\lambda_t-\lambda\lambda_X=F(-\lambda,X;t)\label{gmon}
\ee

\subsection{The second proof}

Let us present the solution of equation of the second order from the
assertion
in the form $Q(X,t)={\rm constant}$. Then for derivatives $\dot X$ (
with the 
help of the theorem of differentiation of  implicit functions) we
obtain:
$$
\dot X=-{Q_t\over Q_X}
$$
Repeating this operation once more we come to the generalized Bateman
equation 
from this assertion (with the obvious exchange  $Q\to\lambda$).

The corollary of the results of last two subsections may be summarised
in:

Proposition (equivalent to the main assertion):

Each equation of second order can be presented in Monge-Bateman form
(\ref{gbat}). If the general solution of an ODE may be presented in
explicit 
form, then  the general solution of (\ref{gmon}) is given by
(\ref{1}).

\section{Multidimensional generalisation}

Suppose we are given a set of $(n-1)$ arbitrary functions $X^i\equiv 
X^i(c^{\alpha};t)$,
each one depending on $2(n-1)$ variables $c^{\alpha}$ and a single
"time" 
variable $t$. All $2(n-1)$ independent variables $c^{\alpha}$ may be
expressed 
implicitly in terms of a system of $2(n-1)$ equations:
\bea
x^i=X^i(c^{\alpha};t),\quad \dot x^i=\frac{\partial X^i(c^{\alpha};t)}
{\partial t}\label{WK}
\eea
 After substitution of these values into explicit expressions for
 second
derivatives we arrive at a system of ordinary differental equations of
second 
order (as described in textbooks):
$$
x^i_{tt}=\frac{\partial^2 X^i}{\partial t^2}(x,x_t;t) 
$$
Now let us consider $2(n-1)$ values $c^{\alpha}$ as a functions of
$(n-1)$
functions $\rho^s\equiv \rho^s (x;t)$. Differentiating the first
equation
of (\ref{WK}) first with respect to $x_j$ and secondly with respect to
$t$ 
 we obtain respectively:
$$
\delta_{ij}=\sum X^i_{\rho^s}\rho^s_{x_j},\quad 0=\sum
X^i_{\rho^s}\rho^s_t+
X^i_t
$$
 From the last equations we obtain immediately:
$$ 
\{X^i_{\rho^s}\}=J^{-1}(\rho;x),\quad X_t=-J^{-1} \rho_t
$$
where $J(\rho;x)$ is the Jacobian matrix. 
Further differentiation of the second equation with respect to $x_i,t$
arguments leads to the result:
$$
\sum X^i_{\rho^s,t}\rho^s_{x_i}=-(J^{-1} \rho_t)_{x_i},\quad
X^i_{tt}+\sum X^i_{\rho^s,t} \rho^s_t=-(J^{-1} \rho_t)t
$$
Eliminating the matrix $\{X^i_{\rho^s,t}\}$ from the last system we
obtain 
finally:
\bea
\tau^s_t-\sum \tau^r \tau^s_{x_r}=X^s_{tt}(x,\tau;t)\label{FM}
\eea
where $\tau=J^{-1} \rho_t$.

In the case of a noninteracting system $X^s_{tt}=0$ (\ref{FM}) goes
over to the 
so called hydrodynamical system introduced and solved by D.B.Fairlie
\cite{DBF}.
In connection with  this, the above implicit solution of the
hydrodynamical 
system arises after treating the constants of motion in the
trajectories of $n$
free moving particles as functions of $n$ arbitrary functions $\rho$.
Thus $n$ 
equations:
$$
x^s=f^s(\rho)t+g^s(\rho)
$$ 
define in implicit form the solution of the $n$ dimensional
Bateman-Monge 
equations. If we choose $f^s=\rho^s$ we reproduce the original form of
the
solution of the hydrodynamical system proposed in \cite{DBF}.

\section{Outlook}

The results contained in formulae (\ref{2}),(\ref{gmon}) and
(\ref{FM}) 
are so simple and clear that they don't demand additional comments. It
is 
possibile that they have been discovered long ago, but the author has
not 
seen them in the literature.

\section*{ Acknowledgements.}

The author gratefully thanks D.B.Fairlie and A.V.Razumov for
discussions in
the process of working on this paper and important comments.

Author is indebted to the Center for Research on Engenering
and Applied Sciences (UAEM, Morelos, Mexico) for its hospitality and
Russian Foundation of Fundamental Researches (RFFI) GRANT N
98-01-00330 for 
partial support.

\end{document}